\documentclass{cimento}

\ifx\pdfoutput\undefined
	\usepackage[dvips]{graphicx}
\else 
	\usepackage[pdftex]{graphicx}
\fi 

\title{Extracting energy from black holes:\ETC \\ ``long'' and ``short'' GRBs and their astrophysical settings}
\author{R.~Ruffini\from{icra}\from{dipfis}\ETC,
M.G.~Bernardini\from{icra}\from{dipfis},
C.L.~Bianco\from{icra}\from{dipfis},
P.~Chardonnet\from{icra}\from{savoie},
F.~Fraschetti\from{icra}\from{trento},
V.~Gurzadyan\from{icra}\from{yerev},
M.~Lattanzi\from{icra}\from{dipfis},
L.~Vitagliano\from{icra}\from{dipfis}
        \atque
S.-S.~Xue\from{icra}\from{dipfis}}

\instlist{
  \inst{icra} ICRA --- International Center for Relativistic Astrophysics.
  \inst{dipfis} Dipartimento di Fisica, Universit\`a di Roma ``La Sapienza'', Piazzale Aldo Moro 5, I-00185 Roma, Italy.
  \inst{savoie} Universit\'e de Savoie, LAPTH - LAPP, BP 110, F­74941 Annecy-le-Vieux Cedex, France.
  \inst{trento} Universit\`a di Trento, Via Sommarive 14, I-38050 Povo (Trento), Italy.
  \inst{yerev} Yerevan Physics Institute, Alikhanian Brothers Street 2, 375036, Yerevan-36, Armenia.
  }
\PACSes{\PACSit{04.70.-s}{Physics of black holes}
\PACSit{98.70.Rz}{gamma-ray sources; gamma-ray bursts}
\PACSit{97.60.Jd}{Neutron stars}
\PACSit{97.60.Bw}{Supernovae}}

\begin{document}

\maketitle

\begin{abstract}
The introduction of the three interpretational paradigms for Gamma-Ray Bursts (GRBs) and recent progress in understanding the X- and $\gamma$ ray luminosity in the afterglow allow us to make assessments about the astrophysical settings of GRBs. In particular, we evidence the distinct possibility that some GRBs occur in a binary system. This subclass of GRBs manifests itself in a ``tryptich'': one component formed by the collapse of a massive star to a black hole, which originates the GRB; a second component by a supernova and a third one by a young neutron star born in the supernova event. Similarly, the understanding of the physics of quantum relativistic processes during the gravitational collapse makes possible precise predictions about the structure of short GRBs.
\end{abstract}

\section{Introduction}

\begin{figure}
\centering
\includegraphics[width=7.3cm,clip]{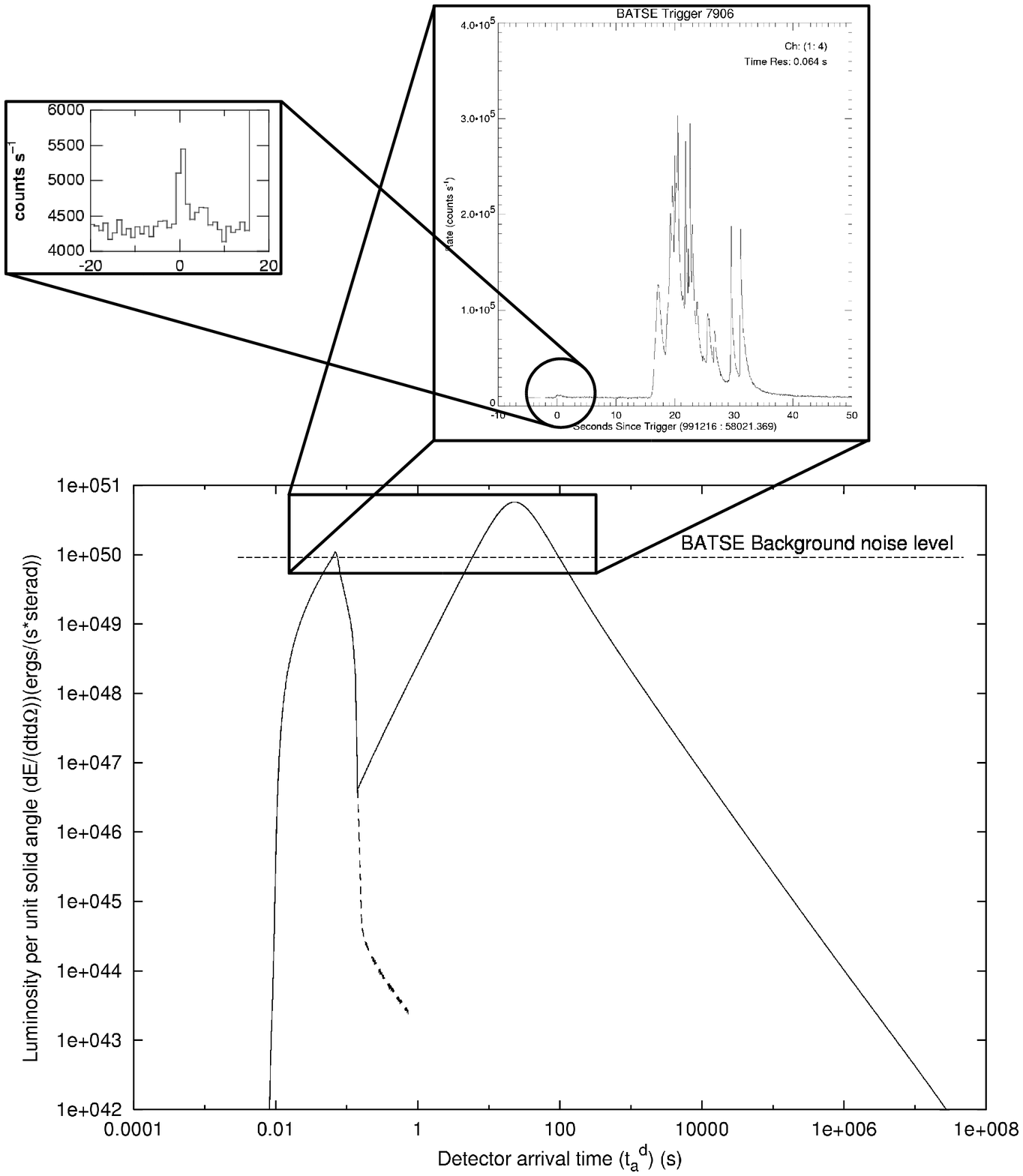}
\includegraphics[width=6cm,clip]{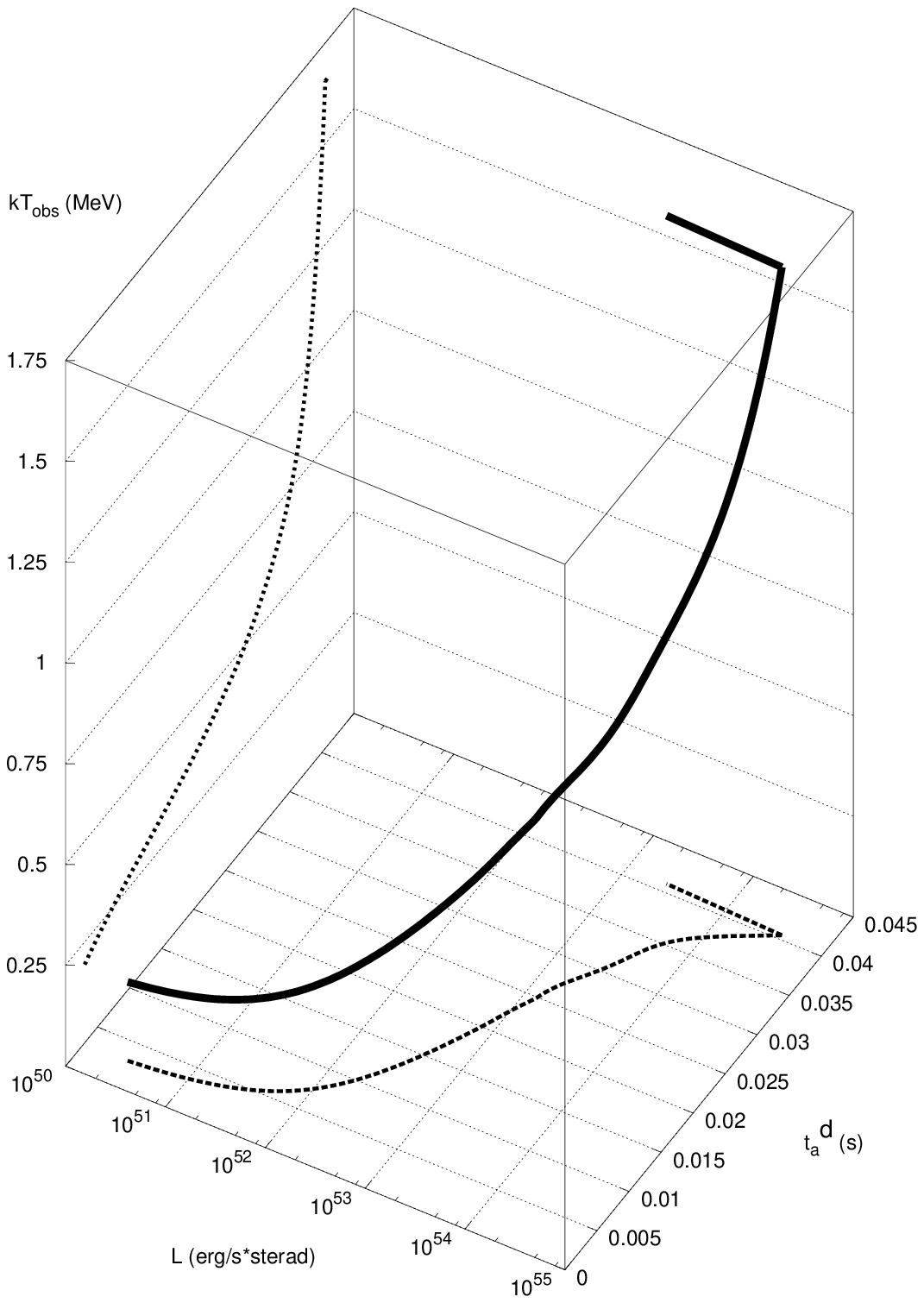}
\caption{{\bf Left:} Our theory applied to GRB 991216 (main panel). It is represented the P-GRB (at arrival time between $0.01$ s and $0.1$ s) and the afterglow starting at $0.1$ seconds. Also represented is the BATSE noise threshold. In the upper right panel is represented the corresponding light curve observed by BATSE. The observed precursor (see the enlargement on the left side) precisely coincides in amplitude and arrival time with the theoretically predicted P-GRB. The main part of the ``prompt radiation'' coincides, both in arrival time and intensity, with the extended emission at the peak of the afterglow (see text). Details in \cite{lett2,Brasile,Brasile2}. {\bf Right:} Predicted observed luminosity and observed spectral hardness of the electromagnetic signal from the gravitational collapse of a collapsing core with $M=10M_{\odot}$, $Q=0.1\sqrt{G}M$ at $z=1$ as functions of the arrival time $t_{a}^d$ at the detector. Details in \cite{rfvx05}.}
\label{ibs}
\label{3d}
\end{figure}

The basic new approach in our model of GRBs has been summarized in three letters \cite{lett1,lett2,lett3} and in a variety of articles, including two extensive review papers \cite{Brasile,Brasile2}. The difference between our model and all other approaches in the literature consists in:\\
a) The spacetime parameterization of the GRB structure \cite{lett1}.\\
b) The identification of two different components in the GRB phenomenon. The first one is the proper-GRB (P-GRB) which is emitted as the optically thick accelerated phase of GRB reaches transparency. The crucial identification of this phenomenon has usually been neglected in the current literature. The second one is the afterglow, originating from the interaction of the baryonic matter, accelerated in the optically thick phase, with the interstellar medium. At variance with the current literature, the commonly called ``prompt radiation'' of the long GRBs is considered a part of the afterglow, being an extended afterglow peak emission (E-APE) \cite{lett2}. The short GRBs are again explained within this unitary model: they correspond to the absence of significant amount of baryonic matter in the optically thick accelerating phase. They coincide with the P-GRBs which in this case are much more prominent, being the afterglow component negligible. See left panel of Fig. \ref{ibs}.\\
c) Although the above considerations ``a'' and ``b'' applies to all GRBs, there are some GRBs which appear to be associated to supernovae. Our model sharply differentiates between the GRB phenomenon and the supernova process \cite{lett3}. It evidences the binary nature of the GRB progenitor and opens the problematic of an induced collapse either by the GRB on the supernova progenitor star or by the supernova on the GRB progenitor star.\\
In addition to these three paradigms, our model advances a specific emission process for the gamma-ray and X-ray afterglows, while optical and radio afterglows (usually observed in the latest phases) are assumed to originate from another emission process \cite{Spectr1,Spectr2}.

On the basis of our GRB model (see \cite{Brasile,Brasile2} and references therein), we have recently fitted GRB 991216, GRB 980425, GRB 030329 and GRB 031203. In all four cases the luminosity and the spectral informations are consistent with isotropic emission and no beaming. Some preliminary results on this last source are presented here. In total analogy with the tryptichs GRB 980425 -- SN1998bw -- URCA-1 and GRB 030329 -- SN2003dh -- URCA-2, we are in presence of a triptych which we call GRB 031203 -- SN2003lw -- URCA-3. We can clearly see from our theoretical prediction on the X- and $\gamma$ ray luminosities (see Fig. \ref{031203}) that there is a late component in the $0.2$--$10$ keV band not connected to the afterglow. This is the emission from URCA-3

\section{On the GRB-Supernova connection}

We first stress some general considerations originating from comparing and contrasting the considered GRB sources:
\begin{enumerate}
\item The value of the $B$ parameter for all sources occurs, as theoretically expected, in the range \cite{Brasile,Brasile2} $10^{-5} \le B \le 10^{-2}$. We have in fact:\\
\begin{tabular}{c|c|c|c|c}
GRB & $B$ & $E_{dya}$ (erg) & $E_{SN}$ (erg) & $\begin{array}{c} L_{URCA}\\ (erg/s/sterad)\end{array}$\\
\hline
991216 & $3.0\times 10^{-3}$ & $4.8 \times 10^{53}$ & --- & ---\\
980425 & $7.0\times 10^{-3}$ & $1.1 \times 10^{48}$ & $\sim 10^{49}$ & $5.2 \times 10^{39}$\\
030329 & $4.8\times 10^{-3}$ & $2.1 \times 10^{52}$ & $\sim 10^{49}$ & $1.9 \times 10^{41}$\\
031203 & $7.4\times 10^{-3}$ & $1.8 \times 10^{50}$ & $\sim 10^{50}$ & $1.0 \times 10^{42}$\\
\end{tabular}
\item The large difference in the GRB energy of the sources simply relates to the electromagnetic energy of the black hole which turns out to be smaller than the critical value. The fact that the theory is valid over $5$ orders of magnitude is indeed very satisfactory.
\item In both sources GRB 980425 and GRB 030329 the associated supernova energies are similar (details in \cite{f03mg10,b03mg10}). The further comparison between the SN luminosity and the GRB intensity is crucial. In the case of GRB 980425 the GRB and the SN energies are comparable, and no dominance of one source over the other can be ascertained. In the case of GRB 030329 the energy of the GRB source is $10^3$ larger than the SN: it is unlikely that the GRB can originate from the SN event.
\end{enumerate}
The above stringent energetic considerations and the fact that GRBs occur also without an observed supernova give further evidence against GRBs originating from supernovae.

\section{URCA-1, URCA-2 and URCA-3}

\begin{figure}
\centering
\includegraphics[width=\hsize,clip]{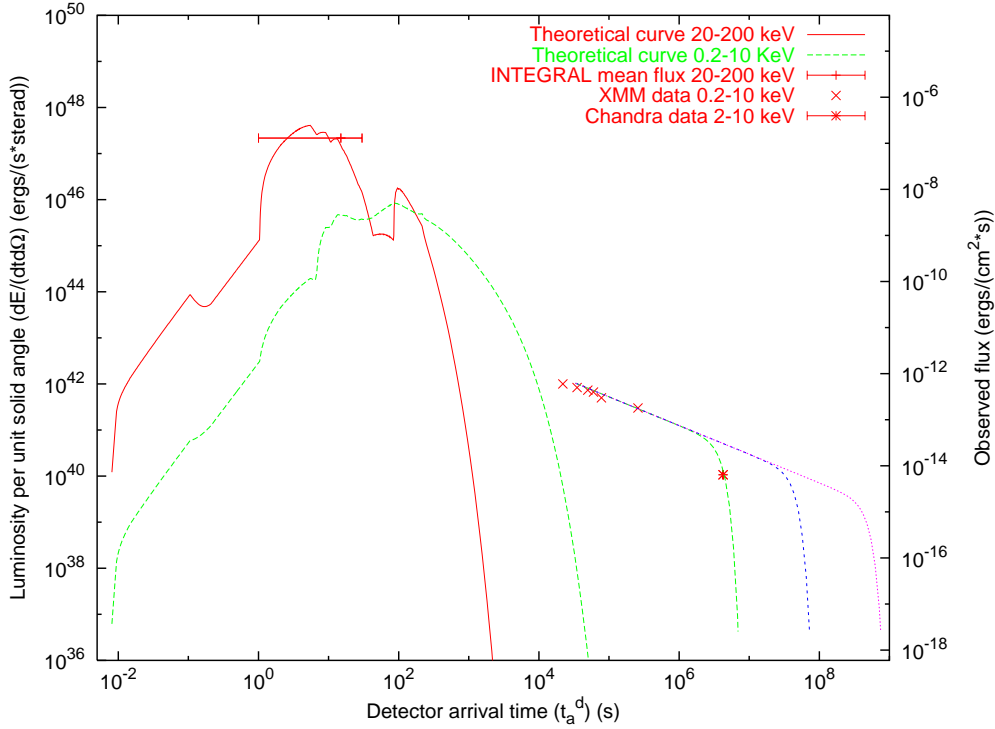}
\caption{Theoretical light curve of GRB 031203 predicted by our model in $\gamma$--rays ($20$--$200$ keV, solid line) and X-rays ($0.2$--$10$ keV, dashed line), together with the observed values (the horizontal bar corresponds to the mean peak flux from \textit{INTEGRAL} \cite{sls04}). The dotted lines correspond to the source URCA-3 compared with the experimental data obtained by \textit{XMM} \cite{wa04} and \textit{Chandra} \cite{fa04}.}
\label{031203}
\end{figure}

We show in Fig. \ref{031203} the X- and $\gamma$ ray luminosities of GRB 031203. We turn then to the exciting search for the nature of URCA-1, URCA-2 and URCA-3. The first possibility is that the URCA sources are related to the black hole originating the GRB phenomenon. In order to probe such an hypothesis, it is enough to find even a single case in which an URCA source occurs in association with a GRB and in absence of an associated supernova. Such a result, theoretically unexpected, would open an entire new problematic in relativistic astrophysics and in the physics of black holes.

On the other hand, if indeed, as we expect, the clear association between URCA sources and the supernovae occurring together with the GRBs will be confirmed, then two other possibilities will be favored. First, an emission from processes occurring into the early phases of the supernova remnant expansion or the very exciting possibility that we are observing a newly born neutron star out of the supernova phenomenon. We shall focus in the following only on this last topic.

The need for a rapid cooling process due to neutrino and anti-neutrino emission in the process of gravitational collapse leading to the formation of a neutron star was considered for the first time by George Gamow and Mario Schoenberg in 1941 \cite{gs41}. It was Gamow who gave to this process the name ``Urca process''. Since then, a systematic analysis of the theory of neutron star cooling was advanced by Tsuruta and collaborators \cite{t64,t79,tc66,t02} and by Canuto \cite{c78}. X-ray observatories such as Einstein (1978-1981), EXOSAT (1983-1986), ROSAT (1990-1998), and the contemporary missions Chandra and XMM-Newton since 1999 had set very embarrassing and stringent upper limits on the neutron star surface temperature in well known historical supernova remnants (see e.g. \cite{r87}). As an example, the neutron star upper limits on the surface temperature in the case of SN 1006 and the Tycho supernova were significantly lower than the temperatures given by standard cooling times (see e.g. \cite{r87}). Much of the theoretical work has been mainly directed, therefore, to find theoretical arguments in order to explain such low surface temperature ($T_s \sim 0.5$--$1.0\times 10^6$ K) --- embarrassingly low, when compared to the initial hot ($\sim 10^{11}$ K) birth of a neutron star in a supernova explosion (see e.g. \cite{r87}). Some relevant steps in this direction have been presented in \cite{vr88,vr91,bl86,lvrpp94,yp04}. The youngest neutron star to be searched for using its thermal emission in this context has been the pulsar PSR J0205+6449 in 3C 58 (see e.g. \cite{yp04}), which is $820$ years old! Recently, evidence for the detection of thermal emission from the crab nebula pulsar was reported in \cite{t05} which is, again, $951$ years old.

In the case of URCA-1, URCA-2 and URCA-3, we may be exploring a totally different regime: the X-ray emission arising from the first months of existence of a neutron star. The reason of approaching first the issue of the thermal emission from the neutron star surface is important, since in principle it may give informations on the equations of state in the core at supranuclear densities, on the detailed mechanism of the formation of the neutron star itself and the related neutrino emission. It is of course possible that the neutron star is initially fast rotating and its early emission is dominated by the magnetospheric emission or by accretion processes from the remnant which would overshadow the thermal emission. In this case a periodic signal related to the neutron star rotational period should in principle be observable if the GRB source is close enough.

Quoting \cite{lvrpp94}: ``The time for a neutron star's center to cool by the direct URCA process to a temperature $T$ has been estimated to be $t = 20 \left[T/\left(10^9 K\right)\right]^{-4}$ s. The direct URCA process and all the exotic cooling mechanisms only occur at supranuclear densities. Matter at subnuclear densities in neutron star crust cools primarily by diffusion of heat to the interior. Thus the surface temperature remains high, in the vicinity of $10^6$ K or more, until the crust's heat reservoir is consumed. After this diffusion time, which is on the order of $1$--$100$ years, the surface temperature abruptly plunges to values below $5 \times 10^5$ K''. This result has been confirmed in \cite{ya01}.

The considerations we have quoted above are developed in the case of spherical symmetry and should keep the minds open to additional factors to be taken into account: 1) the presence of rotation and magnetic field which may affect the thermal conductivity and the structure of the surface, as well as the above mentioned magnetospheric emission; 2) there could be accretion of matter from the expanding nebula; 3) some exciting theoretical possibilities advanced by Dyson on volcanoes on neutron stars \cite{d69} and iron helide on neutron star \cite{d71}; 4) the possibility of piconuclear reactions on neutron star surface discussed in \cite{ls97}.

All the above are just scientific arguments to attract attention on the abrupt fall in luminosity observed in URCA-1, URCA-2 and URCA-3 which is of the greatest scientific interest and further analysis should be followed to check if a similar behavior will be found in future XMM and Chandra observations.

\section{Astrophysical implications}

In addition to these very rich problematics in the field of theoretical physics and theoretical astrophysics, there are also more classical astronomical and astrophysical issues, which will need to be answered if indeed the observations of a young neutron star will be confirmed. An important issue to be addressed will be how the young neutron star can be observed, escaping from being buried under the expelled matter of the collapsing star. A possible explanation can originate from the binary nature of the newly born neutron star: the binary system being formed by the newly formed black hole and the triggered gravitational collapse of a companion evolved star leading, possibly, to a ``kick'' on and ejection of the newly born neutron star. Another possibility, also related to the binary nature of the system, is that the supernova progenitor star has been depleted of its outer layer by dynamic tidal effects.

In addition, there are other topics in which our scenario can open new research directions in fundamental physics and astrophysics:\\
1) The problem of the instability leading to the complete gravitational collapse of a $\sim 10M_\odot$ star needs the introduction of a new critical mass for gravitational collapse, which is quite different from the one for white dwarfs and neutron stars which has been widely discussed in the current literature (see e.g. \cite{gr78}).\\
2) The issue of the trigger of the instability of gravitational collapse induced by the GRB on the progenitor star of the supernova or, vice versa, by the supernova on the progenitor star of the GRB needs accurate timing and the considerations of new relativistic phenomena.\\
3) The general relativistic instability induced on a nearby star by the formation of a black hole needs some very basic new developments in the field of general relativity.

Only a preliminary work exist on this subject (see e.g. \cite{mw}) due to the complexity of the problem: unlike the majority of theoretical work on black holes (which deals mainly with one-body solutions), we have to address here a many-body problem in general relativity. We are starting in these days to reconsider, in this framework, some classic work by Fermi \cite{f21}, Hanni and Ruffini \cite{hr73}, Majumdar \cite{m47}, Papapetrou \cite{p47}, Parker et al. \cite{p73}, Bini et al. \cite{bgr04} which may lead to a new understanding of general relativistic effects relevant to these astrophysical ``triptychs''.

\section{The Short GRBs as cosmological candles}

After concluding the problematic of the long GRBs and their vast astrophysical implications, we briefly turn to the physics of short GRBs \cite{Brasile,Brasile2}. We first recall some progress in the understanding the dynamical phase of collapse, the mass-energy formula and the extraction of blackholic energy which have been motivated by the analysis of the short GRBs \cite{rfvx05}. In this context progress has also been accomplished on establishing an absolute lower limit to the irreducible mass of the black hole as well as on some critical considerations about the relations of general relativity and the second law of thermodynamics \cite{rv02a}. This last issue has been one of the most debated in theoretical physics in the past thirty years \cite{b73,b74,h74,h75,gh77}. Following these conceptual progresses we analyzed the vacuum polarization process around an overcritical collapsing shell \cite{crv02,rv02b}. We evidenced the existence of a separatrix and a dyadosphere trapping surface in the dynamics of the electron-positron plasma generated during the process of gravitational collapse \cite{rvx03b}. We then analyzed, using recent progress in the solution of the Vlasov-Boltzmann-Maxwell system, the oscillation regime in the created electron-positron plasma and their rapid convergence to a thermalized spectrum \cite{rvx03a}. We concluded by making precise predictions for the spectra, the energy fluxes and characteristic time-scales of the radiation for short-bursts \cite{rfvx05} (see right panel in Fig. \ref{3d}).

\cite{ggc03} has found evidence for an exponential cut-off at high energy in the short burst spectra. From the existence of the separatrix introduced in \cite{rvx03a}, we may expect such a cut off (see Fig. \ref{3d}), and are currently comparing and contrasting the observations with our model. If they are in good agreement, this  would lead for the first time to the identification of a process of gravitational collapse and its general relativistic self-closure as seen from an asymptotic observer.

If our theoretical predictions will be confirmed, we will have a powerful tool for cosmological observations: the independent information about luminosity, time-scale and spectrum can uniquely determine the mass, the electromagnetic structure and the distance from the observer of the collapsing core, see \cite{rfvx05}. The short-bursts, in addition to give a detailed information on all general relativistic and relativistic field theory phenomena occurring in the approach to the horizon, may also become the best example of standard candles in cosmology \cite{r03tokyo}.

\section{On the dyadosphere of Kerr-Newman black holes}

An interesting proposal was advanced in 2002 \cite{it02} that the $e^+e^-$ plasma may have a fundamental role as well in the physical process generating jets in the extragalactic radio sources. The concept of dyadosphere originally introduced in Reissner-Nordstr\"{o}m black hole in order to create the $e^+e^-$ plasma relevant for GRBs can also be generalized to the process of vacuum polarization originating in a Kerr-Newman black hole due to magneto-hydrodynamical process of energy extraction (see e.g. \cite{pu01} and references therein). The concept therefore introduced here becomes relevant for both the extraction of rotational and electromagnetic energy from the most general black hole \cite{cr71}.

\acknowledgments
We are thankful to C. Dermer, H. Kleinert, T. Piran, A. Ringwald, R. Sunyaev, L. Titarchuk, J. Wilson and D. Yakovlev for many interesting theoretical discussions, as well as to L. Amati, L.A. Antonelli, E. Costa, F. Frontera, L. Nicastro, E. Pian, L. Piro, M. Tavani and all the BeppoSAX team for assistance in the data analysis, as well as to an anonymous referee for interesting suggestions.

\end{document}